\documentstyle[12pt,epsfig,rotating]{article}
\def\pge{\pagestyle{empty}} \def\pgn{\pagestyle{plain}}
     \def\d{{\rm d}}
\def\spg{\setcounter{page}} 
\def\bd{
\begin{document}} \def\ed{\end{document}}
\def\bmp{\begin{minipage}} \def\emp{\end{minipage}}
\def\bcc{\begin{center}} \def\ecc{\end{center}}     \def\npg{\newpage}
\def\beq{\begin{equation}} \def\eeq{\end{equation}} \def\hph{\hphantom}
\def\be{\begin{equation}} \def\ee{\end{equation}} \def\r#1{$^{[#1]}$}
\def\n{\noindent} \def\ni{\noindent} \def\pa{\parindent} 
\def\hs{\hskip} \def\vs{\vskip} \def\hf{\hfill} \def\ej{\vfill\eject} 
\def\cl{\centerline} \def\ob{\obeylines}  \def\ls{\leftskip}
\def\underbar#1{$\setbox0=\hbox{#1} \dp0=1.5pt \mathsurround=0pt
   \underline{\box0}$}   \def\ub{\underbar}    \def\ul{\underline} 
\def\f{\left} \def\g{\right} \def\e{{\rm e}} \def\o{\over} 
\def\vf{\varphi} \def\pl{\partial} \def\cov{{\rm cov}} \def\ch{{\rm ch}}
\def\la{\langle} \def\ra{\rangle} \def\EE{e$^+$e$^-$}
\def\bitz{\begin{itemize}} \def\eitz{\end{itemize}}
\def\btbl{\begin{tabular}} \def\etbl{\end{tabular}}
\def\btbb{\begin{tabbing}} \def\etbb{\end{tabbing}}
\def\beqar{\begin{eqnarray}} \def\eeqar{\end{eqnarray}}
\def\\{\hfill\break} \def\dit{\item{-}} \def\i{\item} 
\def\bbb{\begin{thebibliography}{9} \itemsep=-1mm}
\def\ebb{\end{thebibliography}} \def\bb{\bibitem}
\def\bpic{\begin{picture}(260,240)} \def\epic{\end{picture}}
\def\akgt{\noindent{Acknowledgements}}
\def\fgn{\noindent{\bf\large\bf Figure captions}}
\bd
\pge
\null{}\vs-2cm
\hskip12cm{\large HZPP-9902}

\hskip12cm{\large Feb. 25, 1999}

\null{}\vs-2cm
\begin{center}
{\Large On the Dominance of Statistical Fluctuation} 

{\Large in the Factorial-Moment Study of Chaos}

{\Large in Low Multiplicity Events of High Energy Collisions 
\footnote{ \ This work is supported in part by the National 
Natural Science Foundation of China (NSFC) \\ \null{}
\hs0.6cm under Grant No.19575021.}}
\vskip0.5cm

{Liu Lianshou \ \ \ \ \ \ \ Fu Jinghua  \ \ \ \ \ \ \ Wu Yuanfang}
\vskip0.0cm

{\small Institute of Particle Physics, Huazhong Normal University, 
Wuhan 430079 China}
\vskip0.0cm

{\small Tel: 027 87673313 \qquad FAX: 027 87662646 
\qquad email: liuls@iopp.ccnu.edu.cn}
\date{ }
\begin{minipage}{125mm}
\vskip 1.5cm
\begin{center}{\Large Abstract}\end{center}
\vskip 0.0cm
\ \ \ \  It is shown using Monte Carlo simulation that for low
multiplicity events the single-event
factorial moments are saturated by the statistical fluctuations. The diverse
of the event-space moments $C_{p,q}$ of single-event moments with the
diminishing of phase space scale, called ``erraticity'', 
observed in experiment can readily be 
reproduced by a flat probability distribution with only statistical 
fluctuations and therefore 
does not indicate the existence of chaos as suggested.  
The possibility of studying chaos in high multiplicity events using 
erraticity analysis is discussed.
\end{minipage}
\end{center}
\vskip 0.1in
{\large PACS number: 13.85 Hd
\vskip0.2cm

\ni

\hskip2.0cm \ Erraticity}

\npg \pgn \spg{2}

Since the experimental observation in 1983 of unexpectedly large local 
fluctuations 
in a single event of very high multiplicity recorded by the JACEE 
collaboration~\cite{JACEE}, the investigation of non-linear phenomena in 
high energy collisions has extracted much attention~\cite{Kittel}. The
anomalous scaling of factorial moments~\cite{BP} averaged over event sample 
called intermittency (or fractal) has been proposed for this purpose.  
Such kind of anomalous scaling has recently been observed successfully
in various experiments~\cite{NA22}\cite{NA27}.

Beside the moments averaged over event sample, the importance of the 
fluctuation of single-event moments inside an event sample has also
been stressed~\cite{CaoHwa}. It is shown that this kind of
fluctuation is related to the chaotic behavior of the system~\cite{CaoHwa}. 
A quantity $\mu_q$ called entropy index has been introduced~\cite{CHPRD}
as an adequate parameter in measuring the chaotic behavior. The positivity of 
entropy index $\mu_q >0$ is proved to be a criterion for chaos~\cite{CHPRE}.
This method has been given the name of ``erraticity analysis''~\cite{err}.

The idea of studying chaos through the fluctuation of single-event moments 
in the event space, in addition to the conventional study of intermittency 
(fractality) through the moments averaged over event sample, is enlightening. 
However, the method of eliminating statistical fluctuations using factorial 
moments, which worked well for the moments averaged over event 
sample~\cite{BP}, can not be simply extended to the case of single-event 
moments, cf. Appendix.  

In this letter we will take this problem into account. We will show,
using Monte Carlo simulation, that the single-event factorial moments are 
saturated by the statistical fluctuations when the multiplicity is low. 
The phenomena observed in experiments~\cite{WSS}, which 
were interpreted as a signal of chaos can readily be reproduced by a flat
probability distribution with only statistical fluctuations. 
The idea~\cite{CaoHwa} of studying chaos in high energy collisions using 
the distribution width of single-event moments in event space is 
meaningful only for high multiplicity events.

In order to study the influence of statistical fluctuations on the chaos-study 
using single-event moments, let us rephrase the formalism~\cite{CaoHwa}
in terms of both the factorial moments and the probability moments parallelly.

The factorial moment $F_q^{(\rm e)}$ and probability moment $C_q^{(\rm e)}$ 
for each event are defined as
\beqar   
F_q^{(\rm e)} &=& \frac
    { \frac{1}{M}\sum\limits_{m=1}^{M} n_m(n_m-1) \cdots (n_m-q+1)} 
    { \f( \frac{1}{M}\sum\limits_{m=1}^{M} n_m\g)^q  },  \\
C_q^{(\rm e)} &=& M^{q-1} \sum_{i=1}^M {\f(p_i^{(\rm e)}\g)}^q 
\eeqar
respectively. 

The moments $F_q^{(\rm e)}$ (or $C_q^{(\rm e)}$) may fluctuate greatly from 
event to event. In a sample consisting of a large number ${\cal N}$ of events, we 
get a distribution 
$P\f(F_q^{(\rm e)}\g)$ $\f({\rm or\ } P\f(C_q^{(\rm e)}\g)\g)$, which is normalized 
to unity. By taking the normalized moments of these distributions in event-space 
defined as
\begin{equation}  
C^{(\rm F)}_{p,q}=\la {F_q^{(e)}}^p\ra \f/  \la F_q^{(e)}\ra^p \g.,
\qquad C^{(\rm C)}_{p,q}=\la {C_q^{(e)}}^p\ra \f/ \la C_q^{(e)}\ra^p \g.,
\end{equation}
we have a quantification of the fluctuation of the spatial patterns, i.e. 
we can investigate the full shape of the distribution and, especially, the 
way it changes with the resolution $\delta=\Delta/M$. The value of $p$ in eqn.(3)
can take any positive real number.  If $C^{(\rm F)}_{p,q}(M)$ 
(or $C^{(\rm C)}_{p,q}(M)$) has a power law behaviour in $M$, i.e.
\begin{equation}   
C^{(\rm F)}_{p,q}(M) \propto M^{\psi^{(\rm F)}_q(p)},
\qquad C^{(\rm C)}_{p,q}(M) \propto M^{\psi^{(\rm C)}_q(p)},
\end{equation}
then the entropy index can be defined as, (For simplicity in notation, the 
superscript (F) and (C) will be omitted in the following.) 
\begin{equation}   
\mu_q=\f.\frac{d}{dp}\psi_q(p)\g|_{p=1}.
\end{equation}

It is easy to see that finite, nonvanishing positive values of $\mu_q$ 
corresponds to wide distribution of $F_q^{(\rm e)}$ (or $C_q^{(\rm e)}$), 
which in turn means unpredictable spatial pattern from event to event. 
By applying the measure to known classical chaotic system, it has been 
shown~\cite{CaoHwa} that $\mu_q$ can be used as a measure of chaos in 
problems where only the spatial patterns can be observed and the positivity 
of $\mu_q$ is a criterion for chaos. 

An alternative way of calculating $\mu_q$~\cite{CHPRE} is to express 
$C_{p,q}$ as
\begin{equation}   
C_{p,q}=\la {\Phi_q^{(e)}}^p\ra,
\end{equation}
in which,
\begin{equation}   
\Phi_q^{(e)}=C_q^{(e)} \f/ \la {C_q^{(e)}}\ra\g. .
\end{equation}
With the definition
\begin{eqnarray}   
\label{e12}
\Sigma_q=\la \Phi_q^{(e)} \ln \Phi_q^{(e)} \ra,
\end{eqnarray}
we can obtain
\begin{equation}   
\mu_q={\frac{\partial \Sigma_q}{\partial \ln M}}
\end{equation}
in the scaling region, i.e. where $\Sigma_q$ exhibits a linear 
dependence on $\ln M$. 

In order to study the influence of statistical fluctuations
let us turn now to the application of this formalism to a simple 
model of dynamical fluctuations --- the random cascading 
$\alpha$-model~\cite{BP}\cite{KXTB} and consider 
both the case of pure probability distribution without any
particle distributed in the space and the case with particle distribution.

In the random cascading $\alpha$-model, the $M$ divisions of 
a phase space region $\Delta$ are made in steps. At the first step, 
it is divided into two equal parts; at the second step,  
each part in the first step is further divided into two equal parts, 
and so on. The steps are repeated until $M= {\Delta Y / \delta y}=2^{\nu}.$
How particles are distributed from step-to-step between the two
parts of a given phase space cell is defined by independent random variable 
$ \omega_{\nu j_{\nu}}$, where $j_{\nu}$ is the position of the sub-cell 
($1\le j_{\nu}\le 2^{\nu}$) and $\nu$ is the number of steps.
It is given by~\cite{KXTB}:
\beq  
\omega_{\nu,2j-1}={1\over 2}(1+\alpha r) \ \ \ ; \ \ \ 
 \omega_{\nu,2j}={1\over 2}(1-\alpha r), \qquad j=1,\dots,2^{\nu-1}
\eeq
where, $r$ is a random number distributed uniformly in the interval
$[-1,1]$. $\alpha$ is a positive number less than unity, which 
determines the region of the random variable $\omega$ and describes 
the strength of dynamical fluctuations in the model. After $\nu$ steps,
the probability in the $m$th window ($m=1,\dots,M$) is 
$p_m=\omega_{1j_1}\omega_{2j_2}\dots \omega_{\nu j_{\nu}}$. 
A certain number $N$ of particles 
will then be put in the $M$ sub-cells according to the multinomial (Bernouli) 
distribution
\beq   
B(n_1,\dots,n_M|p_1,\dots,p_M) = \frac{N!}{n_1!\cdots n_M!} 
p_1^{n_1}\cdots p_M^{n_M} , \qquad \sum_{m=1}^M n_m=N.
\eeq
The factorial moment $F_q^{(e)}$ and probability moment $C_q^{(e)}$ 
in each event can now be calculated according to eqn.(1) and (2) respectively.
Then the moment $C_{p,q}$ and entropy index $\mu_q$ 
of the sample can be obtaioned using eqn.(3) and eqn.(9).

Firstly, let us consider the case of pure probability distribution without
any particle. It can be shown~\cite{FWL} that when the model parameter $\alpha$
is fixed to a constant value there will not be any chaotic behaviour, i.e.
the entropy index $\mu$ will vanish. A positive $\mu$ together with the
anomalous scaling of averaged moment (intermittency) can be obtained
when $\alpha$ is distributed over a certain range. For simplicity let 
$\alpha$ be a random variable having a Gaussian distribution. 
The mean and variance of the Gaussian are both chosen as 0.22.
The results of ln$C_2$, $\ln C_{p,2}$ and
$\Sigma_2$ as function of $\ln M$ are shown in Fig.1($a$). It can be seen
from the figure that both the intermittency (power-law behavior of the averaged 
moment $C_2$) and the chaos (diverse of $C_{p,2}$ for the increasing of $M$)
are reproduced. The intermittency index $\phi_2$ and entropy index $\mu_2$
obtained through linear fit of ln$C_2$ and $\Sigma_2$ vs. ln$M$ (in the
latter case only the last 3 points are used in the fit) are:
\beq   
\phi_2=0.056 \ , \qquad \mu_2=0.016
\eeq
respectively.

Now let us put in a certain number, say 9, of particles according to the
Bernouli distribution, cf. eqn.(11). The results of ln$F_2$, $\ln C_{p,2}$ and
$\Sigma_2$ as function of $\ln M$ are shown in Fig.1($b$). It can be seen
from the figure that both the intermittency and the chaos are survived. 
However, the scale of $C_{p,2}$ increases for an order of magnitude, showing
a much stronger diverse of $C_{p,2}$ with the increasing of $M$. 
The intermittency index $\phi_2$ and entropy index $\mu_2$
obtained  in the same way as  do in the previous no-particle case are
\beq   
\phi_2=0.058 \ , \qquad \mu_2=0.41
\eeq
respectively. 

Comparing eqn.(13) with eqn.(12) we see that the anomalous scaling of the
averaged factorial moment $F_2$ is very similar to that of the averaged
probability moment $C_2$. The corresponding intermittency indices are
almost equal. This means that in the moments averaged over event sample the 
statistical fluctuations are very well eliminated by the factorial moment 
method~\cite{BP}. On the contrary, the entropy index $\mu_2$ for the 
9-particle case is much bigger than that for the pure probability case,
bigger for order of magnitude.
This large ``entropy index'' is clearly comes from statistical fluctuations.

In order to convince us of the dominant role of statistical fluctuations
let us consider the case with perfectly flat probabilty distribution, i.e. 
$p_m=1/M$ ($m=1,\dots,M$), without any dynamical fluctuation. In this case,
calculating with probability moments, both the averaged $C_q$ and the
$C_{p,q}$ are identically equal to unity, and the ln$C_q$, ln$C_{p,q}$ and
$\Sigma_q$ all vanish, i.e. no intermittency, no chaos.

After putting in 9 particles, the averaged factorial moments keep 
constant approximately, independent on $M$, cf. the upper figure of 
Fig.1($c$). (Its value moving down from zero is due to the particular 
normalization used.) Therefore, the intermittency index almost 
vanishes as expected. However, the $C_{p,2}$ diverse strongly, cf. the middle 
figure of Fig.1($c$). The intermittency and entropy indices obtained through
linear fit are
\beq   
\phi_2=0.00046 \ , \qquad \mu_2=0.49
\eeq
respectively. This large value of $\mu_2$ is totally due to statistical
fluctuations. Note that in getting this result the random cascading $\alpha$
model has not been used and therefore the conclusion is model-independent.

The results of flat probability distribution with 9 particles are compared
with those from the NA27 data~\cite{WSS} in Fig.2. It can be seen from the
figure that the fit is very well, which means that the phenomena observed 
in this experiment, 
being dominated by statistical fluctuations, does not indicate the existence 
of chaos.

The particle number 9 used above is about the average multiplicity of the
moderate-energy hadron-hadron collision experiments, such as NA27 and NA22. 
When the number of particle increases, the influence of statistical 
fluctuations decreases. In Fig.3 are shown the ``entropy indices'' $\mu_q$ of 
flat probability distribution with different number $N$ of particles. 
It can be seen that $\mu_q$ decreases with the increasing of $N$ and tends 
to zero when $N$ goes larger and larger. At the upper-right corner  of Fig.3
are plotted the fits of $\mu_q$ vs. $N$ to the empirical formula
\beq
\mu_q = A_q \e ^ {-b_q N^{0.2}}.
\eeq 
The fitting parameters are listed in Table I. 

In Fig.4 are shown the dependence of second order entropy index $\mu_2$ on
the multiplicity for both the flat probability distribution and the 
$\alpha$ model with Gaussian distributed $\alpha$. It can be seen from the
figure that when multiplicity is low, e.g. lower than 50, the two cases
are almost undistinguishable.   It is worthwhile noticing that in this case 
the 'entropy index' $\mu_2$ for flat probability distribution is even higher 
than that for Gaussian distributed $\alpha$. This means that the existence of 
chaotic behavior cannot be revealed using this method. When multiplicty
is high, e.g. higher than 200, the entropy index $\mu_2$ for Gaussian
distributed $\alpha$ becomes greater than that for flat probability distribution
and approaches the physically meaningful entropy index of probability 
distribution (solid line of Fig.4).

It is interesting to try this 
kind of analysis with the central collision data of relativistic heavy ion 
collision experiments, where the multiplicities are
very high. If the entropy indices $\mu_q$ obtained in these experiments turn
out to be considerably bigger than the value for flat probability
distribution, shown in Fig.3 and the full circles of Fig.4, then that will
be a signal of some interesting new physics.

\vs0.5cm
\cl{{\bf Table I} \ Fitting parameters of $\mu_q$ vs. $N$ to eqn.(15)}

\bcc
\begin{tabular}{|c|c|c|}\hline
\ \ $q$ \ \ & \ $A_q$ \ & \ $b_q$ \ \\  \hline
\ \ $2$ \ \ & \ 2.124 \ & \ 3.954 \ \\  \hline
\ \ $3$ \ \ & \ 2.397 \ & \ 3.463 \ \\  \hline
\ \ $4$ \ \ & \ 2.526 \ & \ 3.072 \ \\  \hline
\end{tabular}
\ecc

In summary, it is shown in this letter using Monte Carlo simulation 
that the single-event
factorial moments are saturated by the statistical fluctuations. The diverse
of the event-space moments $C_{p,q}$ of single-event moments with the
diminishing of phase space scale (erraticity) observed in experiment can 
readily be 
reproduced by a flat probability distribution with only statistical 
fluctuations and therefore 
does indicate the existence of chaos as suggested.  
Using ``erraticity analysis'' to study chaos in high energy collisions
is meaningful only for high multiplicity events.

\vs1cm

\ni{\bf Acknowledgement}

The authors are grateful to R. Hwa for helpful discussions.

\newpage
{\pa=0pt
{\ls=2cm{\hs-0.8cm
{\bf Appendix} On the elimination of statistical fluctuations
in the vertically and} 

{\hs1.0cm
horizontally averaged factorial moments} \par} 
\pa=1cm
\vs0.5cm

Firstly, let us recall the elimination of statistical fluctuations in the
vertically averaged FM briefly. The basic assumption was that the statistical 
fluctuations are Poissonian:
$$ Q_n=\int_0^\infty \frac{s^n}{n!} \e^{-s} D(s) \d s ,
\eqno(A1) $$
where $Q_n$ is the multiplicity distribution in the $i$th bin, $D(s)$ the 
dynamical distribution in the same bin. In eqn.(A1) we have used the notation 
$n$ with no subscript to denote the number of particles in the $i$th 
bin~\cite{CHPRE}
The multiplicity distribution $Q_n$ and the dynamical distribution $D(s)$ 
are restricted by the normalization conditions
$$ \sum_{n=0}^\infty Q_n=1, \eqno(A2) $$
$$ \int_0^\infty D(s) \d s=1. \eqno(A3) $$
In order to check whether the equation (A1) is consistent with the conditions
(A2) and (A3), insert (A1) into (A2) 
$$ \sum_{n=0}^\infty Q_n = \int_0^\infty \f(\sum_{n=0}^\infty 
\frac{s^n}{n!} \e^{-s} \g) D(s) \d s .  \eqno(A4)$$
Using the normalization of Poisson distribution
$$ \sum_{n=0}^\infty  \frac{s^{n}}{n!} \e^{-s} =1 ,\eqno(A5)$$
we get
$$ \sum_{n=0}^\infty Q_n = \int_0^\infty D(s) \d s   $$
consistent with eqn.(A2) and (A3).

It should be noticed, however, that the normalization (A5) of Poisson 
distribution
is valid when and only when the summation for $n$ is over {\em all the 
positive integer (and 0)} values of $n$. If some terms in the summation are
missing, or if the summation is ended at some value $n_{\rm max}$ of $n$,
then it will be no longer equal to unity.

Let us consider the elimination of statistical fluctuations in the 2nd order 
FM as example. Consider only the numerator 
$f_2\equiv n^{[2]} \equiv n(n-1)$ in the definition eqn.(1), 
and neglect the denorminator, which is only for normalization. We have
$$ \la f_2\ra_{\rm v} \equiv \la n^{[2]}\ra_{\rm v}
=\frac{1}{\cal N}\sum_{e=1}^{\cal N}  n_i^{(e)}(n_i^{(e)}-1) .
\eqno(A6)$$
The crucial point is that, when the number of event ${\cal N}$ in a sample 
is big, the  bin-multiplicity $n_i^{(e)}$ ($\equiv n$) for fixed $i$ can 
take value from zero to very large 
(constrainted only by energy conservation). Therefore, we have
$$ \la f_2\ra_{\rm v}  = \sum_{n=0}^\infty n(n-1)Q_n = 
\sum_{n=0}^\infty  \int_0^\infty n(n-1)  \frac{s^n}{n!} \e^{-s} D(s) \d s\ 
= \int_0^\infty \sum_{n=0}^\infty  \frac{s^n}{(n-2)!} \e^{-s} D(s) \d s\ .
\eqno(A7) $$
Let $n-2=n'$ and make use of the normalization (A5) of Poisson distribution 
we get
$$ f_2\equiv \la f_2\ra_{\rm v}  = 
\int_0^\infty \f(\sum_{n'=0}^\infty  \frac{s^{n'}}{n'!} \e^{-s}\g) 
s^2 D(s) \d s\ = \int_0^\infty s^2 D(s) \d s \equiv c_2 .
\eqno(A8) $$
This is the well known result ------ (vertically averaged) 
factorial moment (FM) equal to dynamical moment (DM), i.e.
the statistical fluctuations have been eliminated successfully in 
vertically averaging over event sample.

Let us check whether this conclusion is applicable also to  horizontally 
averaging.  There are 2 points that are different to the previous vertically 
averaging case.

{\pa=0pt
{\ls=1.2cm\hs-0.6cm
1) {\em The number $M$ of bins}, unlike the number {$\cal N$} of events, 
{\em cannot be very large}, may be only 1, 2, \dots, or at most some tens 
or hudreds. 

\hs-0.6cm
2) Even more important, {\em the total multiplicity $N^{(e)}$ in a single
event is always fixed}.
\par}

\pa=1cm
We can of course define a {\em bin-multiplicity distribution $Q_n^{(e)}$ of a
single event} as the distribution of bin-multiplicity $n (\equiv n_i^{(e)})$
for fixed $e$ in the $M$ bins, but $Q_n^{(e)}$ is nonzero only for at most 
$M'$ different values of $n$, where $M'\leq M$.
$$ Q_n^{(e)} \neq 0 \ \ {\rm only \ for} \ \ n=n_1, n_2, \dots, n_{M'} . $$
Among the $M'$ integers $n_1, n_2, \dots, n_{M'}$,
there is certainly a maximum one $n_{\rm max}$.
$$ Q_n^{(e)} = 0 \ \ {\rm when} \ \ n>n_{\rm max} . \eqno(A9)$$
Therefore, for the single event bin-multiplicity distribution,
$$ \sum_{n=0}^\infty Q_n^{(e)} = \int_0^\infty \f(\sum_{n=0}^{n_{\rm max}} 
\frac{s^n}{n!} \e^{-s} \g) D(s) \d s , $$
Since the summation over $n$ in the round bracket ends at a finite number
$n_{\rm max}$ and has only $M'$ terms, it does not equal to unity and
$$ \sum_{n=0}^\infty Q_n^{(e)} \neq \int_0^\infty D(s) \d s . \eqno(A10)$$
This means that eqn.(A1) is inconsistent with eqn.(A2) and (A3) for $Q_n^{(e)}$.
In other words, {\em for the single-event bin-multiplicity distribution 
$Q_n^{(e)}$ normalized as eqn.(A2) no normalizable $D(s)$ which satisfies 
eqn.(A1) can exist}. 
Therefore, the proof for the elimination of statistical fluctuations
is no longer valid in the horizontal averaging case.

State alternatively, if we write 
$$ \la f_2\ra_{\rm h}  = \sum_{n=1}^\infty Q_n^{(e)} n(n-1) ,  \eqno(A12) $$
then due to eqn.(A9) the summation in $n$ doesnot really extend to infinity.
If we neglect the non-existence of $D(s)$ for a while and try to do the 
calculation similar to eqn.(A7),(A8), we will get
\begin{eqnarray*}
\la f_2\ra_{\rm h}  &=& \sum_{n=1}^{n_{\rm max}} n(n-1) Q_n^{(e)} \
= \sum_{n=1}^{n_{\rm max}} \int_0^\infty n(n-1)  
\frac{s^n}{n!} \e^{-s} D(s) \d s \\
&=& \int_0^\infty \f( \sum_{n=1}^{n_{\rm max}}  
\frac{s^n}{(n-2)!} \e^{-s}\g) D(s) \d s 
= \int_0^\infty \f(\sum_{n'=1}^{n_{\rm max}-2} \frac{s^{n'}}{n'!} \e^{-s}\g) 
s^2 D(s) \d s . \\
\end{eqnarray*}
The normalization condition (A5) cannot be used here, and therefore
$$ \la f_2\ra_{\rm h} \ne \int_0^\infty s^2 D(s) \d s . \eqno(A13) $$

\newpage
\def\Journal#1#2#3#4{{#1} {\bf #2} (#3) #4}
\def\NCA{\em Nuovo Cimento} \def\NIM{\em Nucl. Instrum. Methods}
\def\NIMA{{\em Nucl. Instrum. Methods} A} \def\NPB{{\em Nucl. Phys.} B}
\def\PLB{{\em Phys. Lett.}  B} \def\PRL{\em Phys. Rev. Lett.}
\def\PRD{{\em Phys. Rev.} D} \def\ZPC{{\em Z. Phys.} C}
\def\PRE{{\em Phys. Rev.} E} \def\PRC{{\em Phys. Rev.} C} 

\newpage
\vskip1cm
\ni
{\Large\bf Figure Captions}
\vskip0.8cm

{\pa=0pt{\ls=15mm\rightskip15mm
\hs-15mm
{\bf Fig.1} \ Logarithm of averaged moment ln$C_2$ (ln$F_2$), ln$C_{p,2}$ 
and $\Sigma_2$ as function of ln$M$ for ($a$) the probability moments of
$\alpha$ model with Gaussian-distributed $\alpha$; ($b$) the factorial
moments of the same model with 9 particles put in; ($c$) the factorial
moments of a fat probability distribution with 9 particles put in. 
solid lines are linear fit. Dashed lines are for guiding the eye.
\par}}
\vskip0.5cm

{\pa=0pt{\ls=15mm\rightskip15mm
\hs-15mm
{\bf Fig.2} \ ln$C_{p,q}$ ($q=2$--4) vs. ln$M$ for a flat distribution
with 9 particles put in (full squares) compared with the experimental 
results of NA27 data (open circles). Data taken from Ref.\cite{WSS}.
\par}}
\vskip0.5cm

{\pa=0pt{\ls=15mm\rightskip15mm
\hs-15mm
{\bf Fig.3} \ The dependence of the entropy indices $\mu_q$ on the 
number of particles. The solid lines are the results from the empirical
formula (15) and Table I.
\par}}
\vskip0.5cm

{\pa=0pt{\ls=15mm\rightskip15mm
\hs-15mm
{\bf Fig.4} \ The dependence on the number of particles of the entropy 
indices $\mu_2$ calculated from factorial moments. Full circles ------ flat
probability distribution with only statistical fluctuations; open triangles
------ Gaussian distributed $\alpha$. The solid line is the result 
from probability moments with Gaussian distributed $\alpha$.
The dashed lines are for guiding the eye.
\par}}

\newpage
\baselineskip 0.18in

\begin{picture} (260,240) 
\put(-95,-420)   
{\epsfig{file=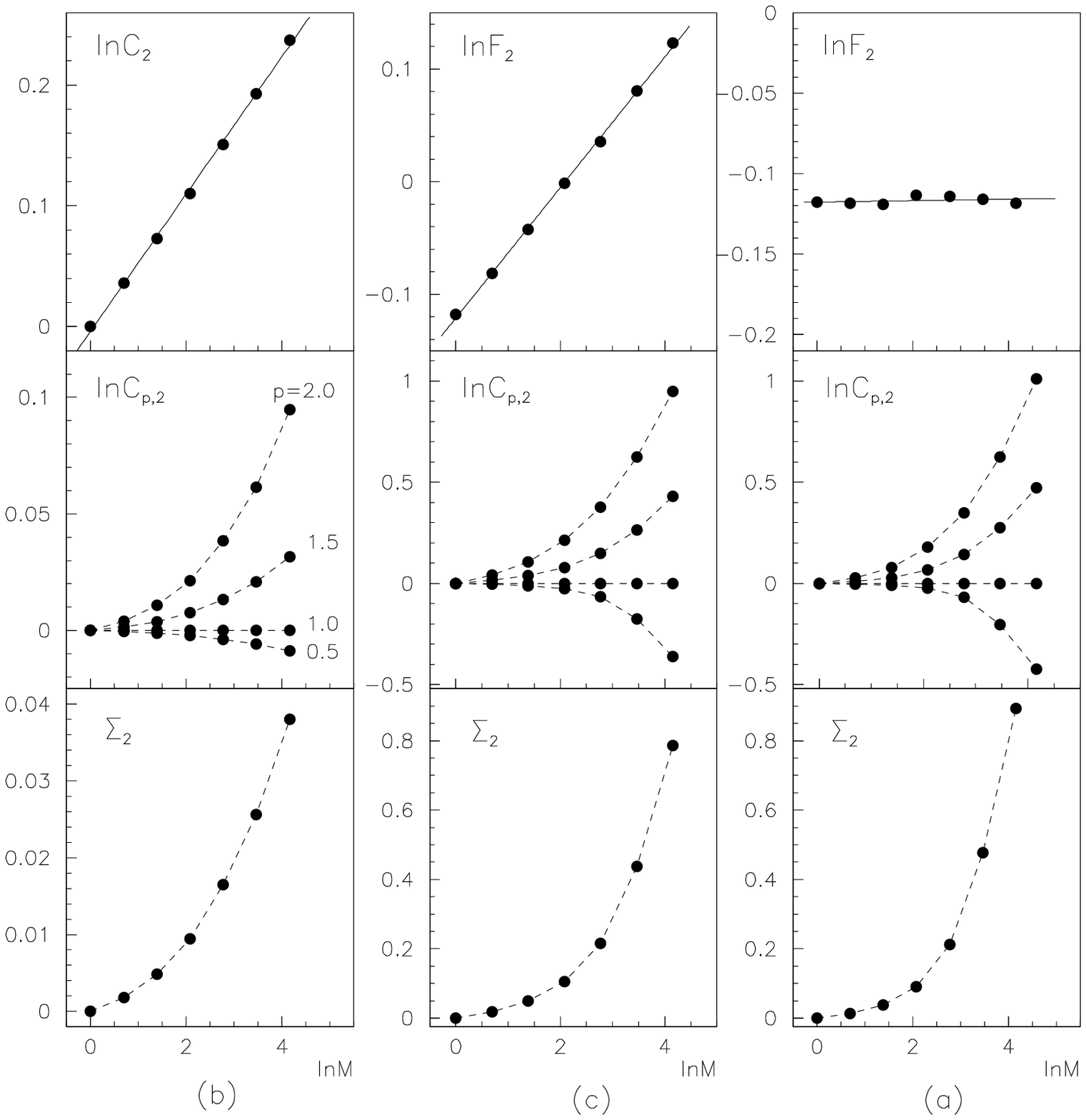,bbllx=0cm,bblly=0cm,
           bburx=8cm,bbury=6cm}}  
\end{picture}
 \vs11.0cm

{\pa=0pt{\ls=15mm\rightskip15mm
\hs-15mm
{\bf Fig.1} \ Logarithm of averaged moment ln$C_2$ (ln$F_2$) and ln$C_{p,2}$ 
as function of ln$M$ for ($a$) the probability moments of
$\alpha$ model with Gaussian-distributed $\alpha$; ($b$) the factorial
moments of the same model with 9 particles put in; ($c$) the factorial
moments of a fat probability distribution with 9 particles put in. 
Solid lines are linear fit. Dashed lines are for guiding the eye.
\par}}

\begin{picture} (260,240) 
\put(-75,-290)  
{\epsfig{file=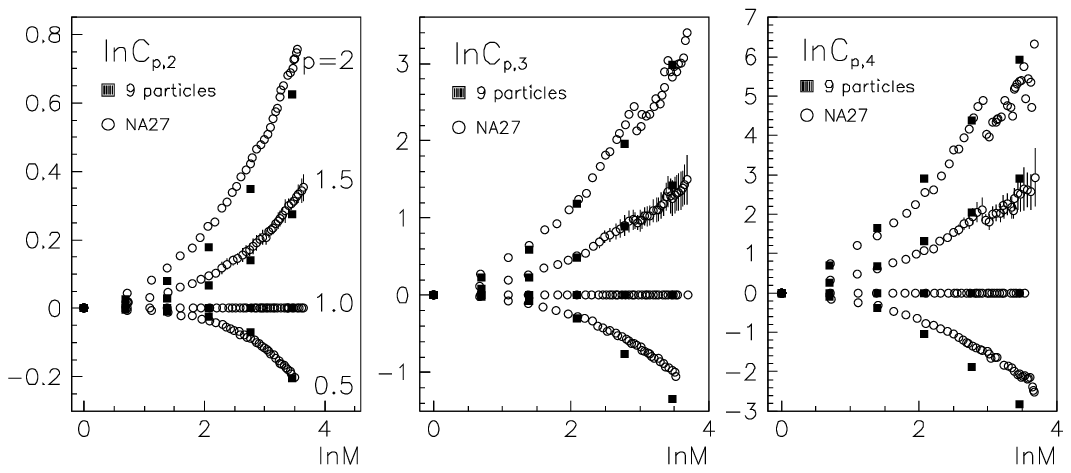,bbllx=0cm,bblly=0cm,
           bburx=8cm,bbury=6cm}}  
\end{picture}
\vskip-1.5cm

{\pa=0pt{\ls=25mm\rightskip15mm
\hs-15mm
{\bf Fig.2} \ ln$C_{p,q}$ ($q=2$--4) vs. ln$M$ for a flat distribution
with 9 particles put in (full squares) compared with the experimental 
results of NA27 data (open circles). Data taken from Ref.\cite{WSS}.
\par}}

\begin{picture} (260,240) 
\put(-75,-390)   
{\epsfig{file=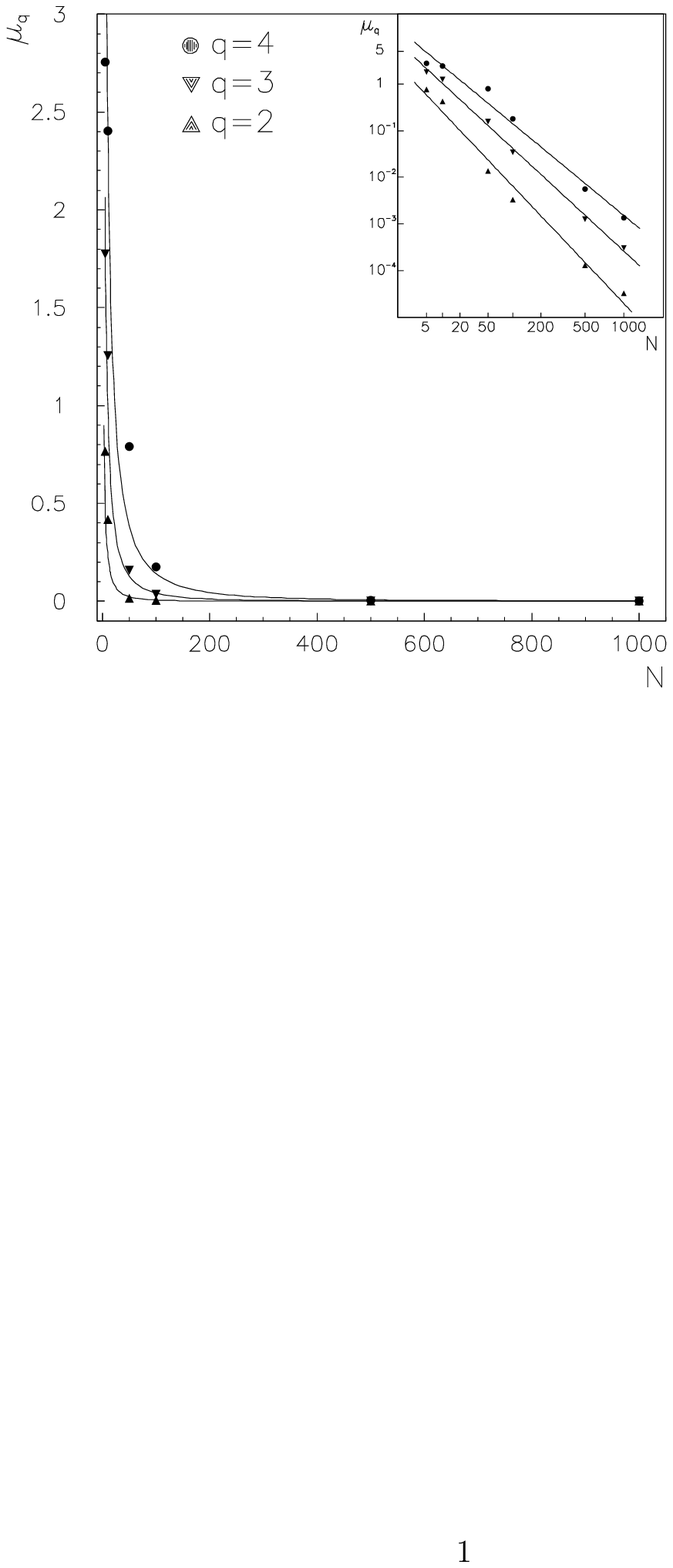,bbllx=0cm,bblly=0cm,
           bburx=8cm,bbury=6cm}}  
\end{picture}
\vskip2.0cm

{\pa=0pt{\ls=25mm\rightskip15mm
\hs-15mm
{\bf Fig.3} \ The dependence of the entropy indices $\mu_q$ on the 
number of particles. The solid lines are the results from the empirical
formula (15) and Table I.
\par}}

\begin{picture} (260,240) 
\put(-75,-450)   
{\epsfig{file=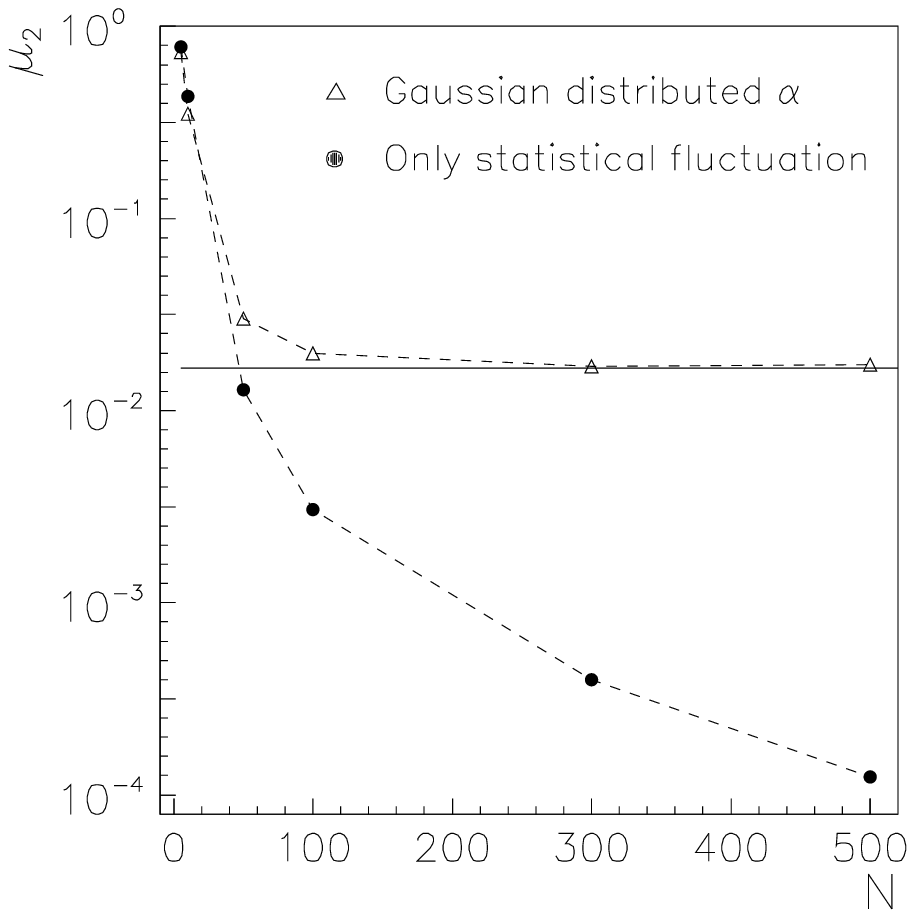,bbllx=0cm,bblly=0cm,
           bburx=8cm,bbury=6cm}}  
\end{picture}
\vskip7.5cm

{\pa=0pt{\ls=25mm\rightskip15mm
\hs-15mm
{\bf Fig.4} \ The dependence on the number of particles of the entropy 
indices $\mu_2$ calculated from factorial moments. Full circles ------ flat
probability distribution with only statistical fluctuations; open triangles
------ Gaussian distributed $\alpha$. The solid line is the result 
from probability moments with Gaussian distributed $\alpha$.
The dashed lines are for guiding the eye.
\par}}

\ed